\documentclass[a4paper]{aa}

\usepackage{graphicx}
\usepackage{graphics}
\usepackage{epsfig}
\usepackage{bm}
\usepackage{amsmath,amssymb,amsfonts}

\usepackage[OT1]{fontenc}
\usepackage[english]{babel}
\usepackage{natbib}

\newcommand{\etal}{et al. }
\newcommand{\dd}{\mathrm{d}}

\newcommand{\mat}{\mathrm{m}}

\newcommand{\baryons}{\mathrm{b}}
\newcommand{\cdm}{\mathrm{cdm}}
\newcommand{\de}{\mathrm{de}}

\newcommand{\Om}{\Omega_\mat}
\newcommand{\Ocdm}{\Omega_\mathrm{cdm}}
\newcommand{\On}{\Omega_\nu}
\newcommand{\Ob}{\Omega_\mathrm{b}}

\newcommand{\hh}{h^2}
\newcommand{\Neff}{N_\mathrm{eff}}
\newcommand{\ns}{n_\mathrm{s}}

\newcommand{\lin}{{\small\mathrm{lin}}}
\newcommand{\nonlin}{{\small\mathrm{nl}}}
\newcommand{\Map}[1]{\left<M^2_\mathrm{ap}( #1 )\right>}
\newcommand{\summnu} {\sum m_\nu}

\begin{document}

 \title{CFHTLS\thanks{Based on
   observations obtained with MegaPrime/MegaCam, a joint project of CFHT
   and CEA/DAPNIA, at the Canada-France-Hawaii Telescope (CFHT) which is
   operated by the National Research Council (NRC) of Canada, the Institut
   National des Sciences de l'Univers of the Centre National de la Recherche
   Scientifique (CNRS) of France, and the University of Hawaii. This work
   is based in part on data products produced at TERAPIX and the Canadian
   Astronomy Data Centre as part of the Canada-France-Hawaii Telescope
   Legacy Survey, a collaborative project of NRC and CNRS.} weak-lensing
   constraints on the neutrino masses}

 \author{
   I. Tereno\inst{1,2}
   \and
   C. Schimd\inst{3,4}
   \and
   J.-P. Uzan\inst{2}
   \and
   M. Kilbinger\inst{2}
   \and
   F.~H. Vincent\inst{2}
   \and
   L. Fu\inst{5,2,6}
   }

 \offprints{{\tt tereno@astro.uni-bonn.de}}

 \institute{
 Argelander-Institut f\"{u}r Astronomie, Auf dem H\"{u}gel 71,
 53121 Bonn, Germany \and
 Institut d'Astrophysique de Paris, CNRS UMR 7095 \& UPMC, 98 bis bd Arago, 75014 Paris, France \and
 Laboratoire d'Astrophysique de Marseille, 38 rue Joliot-Curie, 13388 Marseille, France \and
 Universit\'{e} de Provence -- Aix - Marseille I , Marseille, France \and
 INAF - Osservatorio Astronomico di Capodimonte, via Moiariello 16, 80131
 Napoli, Italy \and
Shanghai Key Lab for Astrophysics, Shanghai Normal University,
200234 Shanghai, P. R. China.
}

   \date{Received 3 October 2008 ; accepted 7 February 2009}


\abstract
{Oscillation experiments yield strong evidence that at least some neutrinos
  are massive. As a hot dark-matter component, massive neutrinos shuld modify
  the expansion history of the Universe as well as the evolution of
  cosmological perturbations, in a different way from cold dark matter or dark energy.}
{We use the latest release of CFHTLS cosmic-shear data to constrain the
sum of the masses $\sum m_\nu$ of neutrinos, assuming three degenerate mass states.
We also consider a joint analysis including other cosmological
observables, notably CMB anisotropies, baryonic acoustic oscillations, and distance
modulus from Type Ia supernovae.}
{Combining CAMB with a lensing code, we compute the aperture mass
variance using a suitable recipe to deal with matter perturbations
in the non-linear regime. The statistical analysis is performed by
sampling an 8-dimensional likelihood on a regular grid as well as
using the importance sampling technique.}
{We obtain the first constraint on neutrino masses based on cosmic-shear data,
  and combine CFHTLS with WMAP, SDSS, 2dFGRS, Gold-set, and SNLS data. The
  joint analysis yields 0.03~eV $ < \sum m_\nu < 0.54$~eV at the 95\%
  confidence level. The preference for massive neutrinos vanishes when 
  systematics are included.}
{}

   \maketitle


\section{Introduction}\label{sec1}
The construction of a cosmological model \citep{Bondi1960, Ellis1971},
must take into account any progress in the
understanding of the laws of physics. To date, the reference model
$\Lambda$CDM is based on the standard model of particle physics,
general relativity, and some additional hypothesis about the symmetries
of the background spacetime (the Copernican principle) to which it
is mandatory to add the two still unknown components of cold dark matter
(CDM) and cosmological constant ($\Lambda$).

We know now that it is imperative to include the effects of massive
neutrinos, which behave as a warm or hot dark matter component depending
on their mass. Strong evidence of such particles, which have not yet been 
included in
the standard model of particle physics but whose existance is supported by
e.g. grand-unified
theories \cite{GUT}, emerges from experimental results on oscillations
of atmospheric, solar, and accelerator- or reactor-produced neutrinos, 
such as Super-Kamiokande, K2K, MINOS, KamLAND \cite[see][for a review]{NuExpReview}.
These results indicate a mixing of the three known neutrinos species by
non-vanishing squared mass differences between the mass eigenstates
and the non-vanishing corresponding mixing angles. The most recent
results are $\Delta m_{12}^2\sim 8\times 10^{-5}$~eV$^2$, $\Delta m_{23}^2\sim
2\times 10^{-3}$~eV$^2$, and mixing angles of $\theta_{12} \sim 30^\circ$,
$\theta_{23}\sim 45^\circ$, and $\theta_{13} \lesssim 11^\circ$.
These values yield a lower bound on the heavier mass of order $0.06 \,(0.1)$~eV
for normal (inverted) hierarchy, but cannot give the absolute mass scale.
Forthcoming experiments based on other mechanisms, such as SuperNEMO
(conceived to detect the neutrinoless double $\beta$ decay) and KATRIN
(designed to probe the tritium $\beta$ decay) are designed to measure
directly the \emph{absolute} mass of the electron neutrino with
a sensitivity of 0.05 and 0.2~eV, respectively.

Cosmological observations are compelling in providing an
independent way to probe the absolute neutrino mass. The effect of
light massive fermions in cosmology is well understood
\cite[see][for a review]{NuCosmoReview}. The standard cosmological
model predicts the existence of relic neutrinos that decouple
from the primeval plasma while ultra-relativistic and produce a
cosmic neutrino background, which cannot yet be detected directly.
Their contribution to the total radiation energy density is a
fraction $7/8(T_\nu/T_\gamma)^4\Neff$ of the contribution of
photons, $T_\nu$ and $T_\gamma$ being the
temperatures of the neutrino and the photon backgrounds, respectively. The
effective number of neutrinos is constrained to be $\Neff=3.04$, if
there are no extra relativistic degrees of freedom besides three active
neutrinos \cite{Neff}, and has an impact mainly on the primordial
nucleosynthesis and the time of matter-radiation equality.
Analyses of light element abundances \cite{NeffBBN} and of the
time of equality \cite{wmap5komatsu} produce results that agree with this
value.

When neutrinos become non-relativistic, their velocities begin to decrease from
$c$ as $T_\nu(t)$, due to momentum conservation. Assuming that the mass states are
degenerate ($m_1=m_2=m_3\equiv m_\nu$), the velocity is given by
$v_\mathrm{th}=3k_BT_\nu(t)/m_\nu c$ or, since $T_\nu(z)\simeq 1.9(1+z)$~K, by
\begin{equation}\label{eq:vth}
v_\mathrm{th}= 150\,(1+z)(m_\nu/1\,\mathrm{eV})^{-1}~\mathrm{km/s}.
\end{equation}
The transition occurs at
\begin{equation}\label{eq:ztrans}
1+z_\mathrm{tr}= 2\times 10^3\left(\frac{m_\nu}{1\,\mathrm{eV}}\right).
\end{equation}
For masses smaller than $\sim$ 0.6 eV, the transition occurs well into the
matter-dominated era and after recombination. 
Therefore, massive neutrinos affect the anisotropies of the CMB temperature
only through the background evolution. 
However CMB data strongly constrain other cosmological parameters that are
degenerate with the neutrino mass, and is thus invaluable to joint analyses. 
The result would slightly differ for non-degenerate masses, 
for the normal or inverted hierarchies, but at a negligible level for a total mass above 0.2 eV \cite{lesgourgueshepph}.

After the transition, neutrino-density perturbations evolve in a similar way to
CDM on scales larger than a free-streaming length, and are damped
on smaller scales. The comoving free-streaming length is defined analogously
to a Jeans length, by replacing the sound of speed by $v_\mathrm{th}$, i.e., 
$\lambda_\mathrm{fs}(t)=2\pi\sqrt{2/3}\,v_\mathrm{th}(t)/[a(t)H(t)]$.
It decreases from the time of transition onwards as, 
\begin{equation}\label{eq:kfs}
\lambda_\mathrm{fs} = 2 \pi \times 1.2 (1+z)^{1/2} (m_\nu/1\,\mathrm{eV})^{-1}(\Om
h^2)^{-1/2} \,\,\mathrm{Mpc},
\end{equation}
which is valid at high redshift. The maximum occurs at the transition and is given by,
\begin{equation}\label{eq:kfsmax}
\lambda^*_\mathrm{fs}= 2\pi \times 55 (\Om h^2)^{-1/2}\left( \frac{m_\nu}{1\,\mathrm{eV}}\right)^{-1/2}h^{-1} \mathrm{Mpc}.
\end{equation}
The density perturbations of the other components, in particular CDM, are
affected by the presence of massive neutrinos as a consequence of the change in the background evolution, the change in the time of equality, and
feedback from the neutrino perturbations.
Roughly speaking, the effect of $m_\nu=1\mathrm{eV}$ neutrinos in the power
spectrum is similar to the CDM effect on large scales $k=2\pi/\lambda < k_\mathrm{fs}(z_\mathrm{tr}) \simeq 0.01\,
\mathrm{Mpc}^{-1}$; a scale-dependent suppression of the power spectrum amplitude 
at intermediate scales $0.01\,\mathrm{Mpc}^{-1}\lesssim k < k_\mathrm{fs}(z=0)\simeq 0.5\,\mathrm{Mpc}^{-1}$;
and a scale-independent suppression of amplitude at small scales $k \gtrsim 0.5\,\mathrm{Mpc}^{-1}$.

Various combinations of cosmological data have been used to constrain
the neutrino mass. 
Using CMB data alone, the WMAP 5-year analysis obtaied an upper 95\% confidence
limit on the sum of the neutrinos masses of 1.3~eV, for a $\Lambda$CDM
cosmology \cite{wmap5komatsu}. By adding galaxy clustering data to CMB data and marginalizing
over the galactic bias, the upper limit decreases to around  0.7~eV - 1.0~eV,
depending on the datasets used \citep[e.g.,][]{Tegmark06, Fogli08}.
If instead, distance indicators (supernovae and
baryon acoustic oscillations) were combined with CMB, stronger
constraints were obtained, e.g., $\summnu < 0.61$~eV \cite{wmap5komatsu}.
When all of these probes were used together, the result improved to around
0.4~eV - 0.5~eV \citep[e.g.,][]{Goobar, Kristiansen07}. Similar results
were obtained using only CMB and galaxy clustering data but assuming that the
bias was known \cite{Mctavish06}, or measuring it independently
\cite{biasnu}. By applying the latter technique to WMAP 5-year data and new
galactic bias constraints, an improved value of
$\summnu < 0.28$~eV was obtained \cite{Debernardis}.
Lyman-$\alpha$ data was also added to different combinations of probes,
producing results of around 0.15~eV - 0.3~eV, including the tighest constraint to
date of $\summnu < 0.17$~eV \cite{Seljaklyalpha}.

In this paper we present the first constraints on neutrino masses
obtained with cosmic shear data. 
The possibility of using the effect of weak gravitational lensing
of background galaxies by large-scale structures, or cosmic shear
\cite[see][for a review]{physrepwl}, to constrain neutrino properties was
first studied in Cooray (1999). Since then, various forecasts have been made
and future cosmic shear surveys in combinations
with redshift information and CMB data from Planck are
expected to measure the sum of neutrino masses with an accuracy of
$~0.05$~eV, using shear tomography \cite{Hannestad} or the full
3D shear field \cite{nu3d}.
In the present work, we use the latest release of the
CFHTLS wide survey, CFHTLS-T0003, 
where two-point angular correlations of the cosmic shear field were detected
 on scales ranging from 1\arcmin \, to 4\degr \, \cite{Fu}.
The signal was measured on source galaxies with mean redshift
$z_\mathrm{m}=0.92$, implying it is produced mainly by the dark matter
distribution at $z \simeq 0.4$. 
The measured range therefore probes
the non-linear matter power spectrum on comoving scales
of $0.08 < k < 10\,h$~Mpc$^{-1}$, reaching the quasi-linear regime
 of largest scale corresponding to a physical size of approxiamtely $1\,h^{-1}$~Gpc. 
This largest scale probed is still one order of magnitude smaller than 
 the largest free-streaming scale, $k^*_\mathrm{fs}\simeq 0.01\,
\mathrm{Mpc}^{-1}$, computed from Eq.~(\ref{eq:kfsmax}). The data probes thus 
both the scale-independent and the
scale-dependent suppression effects of the sub-free-streaming regime.

\begin{figure*}[tb]
\centering
\hspace{-1cm}
\includegraphics[width=7cm]{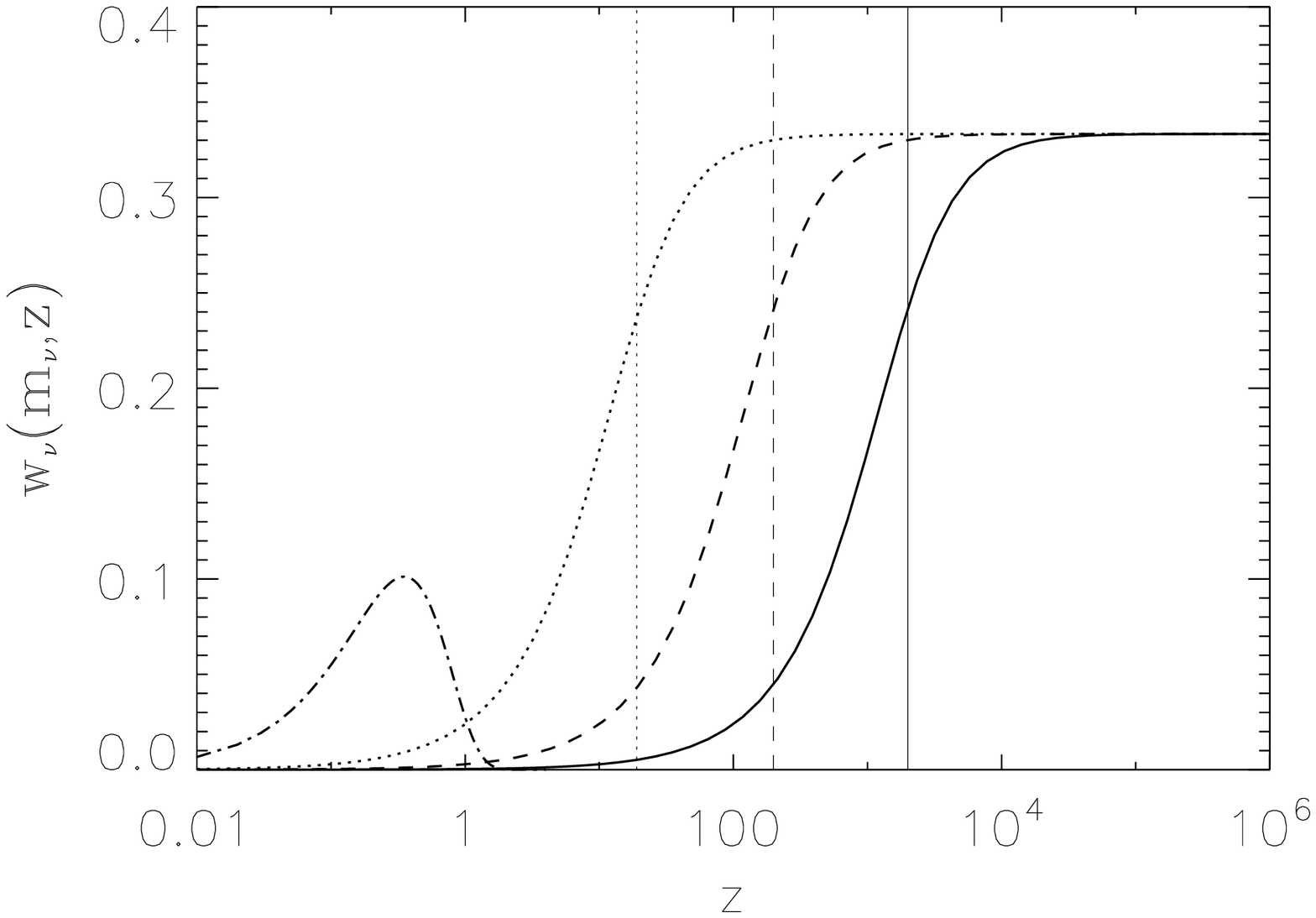}
\hspace{1cm}
\includegraphics[width=7cm]{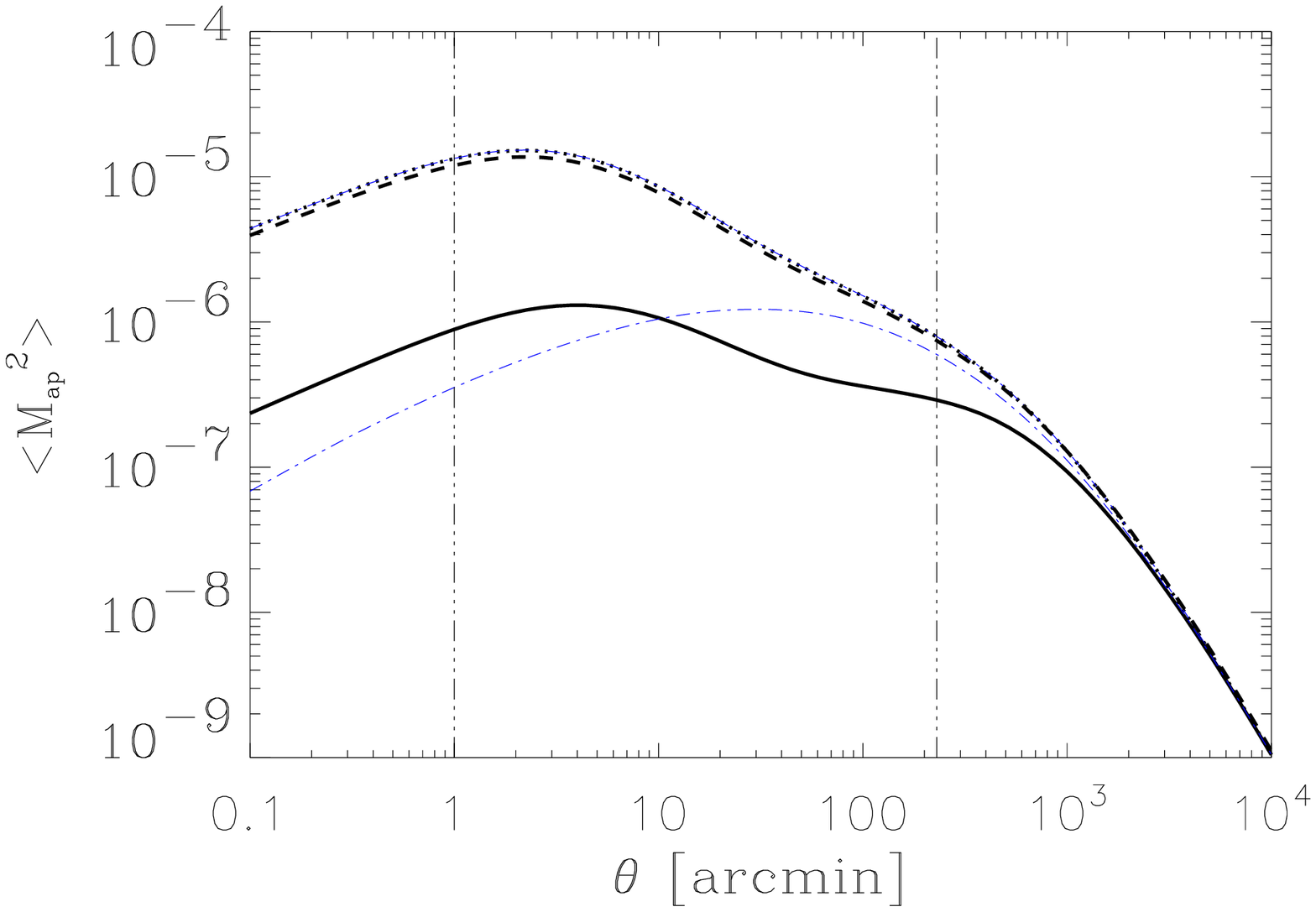}
\vspace{0.4cm}
\caption{ {\it Left:} Equation-of-state of fermionic massive
particles as function of redshift $z$ for $m=1.0$~eV (solid line),
0.1~eV (dashed) and 0.01~eV (dotted). The transition redshift,
Eq.~(\ref{eq:ztrans}), is indicated in each case. The
dashed-dotted curve shows the lensing efficiency window (arbitrary
normalization). {\it Right:} Aperture-mass variance $\Map{\theta}$
with massive and massless neutrinos as function of angular scale.
The massless neutrinos model (blue, thin lines) is shown in
both linear and non-linear ({\sc halofit}) approximations. This 
 indicates that the CFHTLS-T0003 data range (defined by the
vertical lines) lies mostly in the non-linear regime, where the
linear-approximation curve is much lower, and reaches the 
quasi-linear regime. Black, thick curves show three models of
massive neutrinos (same masses and line types as in the left
panel) with identical total mass densities ($\Om\hh=0.14$). The models for the
two lighter neutrino cases are indistinguishable,
while the case $m_\nu=1.0$~eV produces a significantly lower signal at small
scales.} \label{fig:wmz+deltaMap2}
\end{figure*}

We proceed by presenting in Sect. 2 the methodology followed to compute
the spectra. In the same section we introduce the statistical analysis
performed and the datasets used. Cosmological constraints are obtained and
discussed in Sect. 3 and our conclusions are presented in Sect. 4.

\section{Methodology}\label{sec2}

\subsection{Implementing massive neutrinos in cosmic-shear 2-point correlation functions}\label{sec2a}

We perform a statistical analysis of a mixed dark-matter scenario by
considering a 8+1 dimensional parameter space (see Sect.
\ref{sec2b}). For each sampled model of this parameter space the
cosmic-shear power spectrum $P_\gamma(\ell)$ is computed by integrating the
non-linear \emph{total} matter power spectrum  $P_\mat^{\mathrm nl}$ along the
line-of-sight, up to the limiting redshift of the survey
$z_\mathrm{lim}$, to account for the distribution of source galaxies
$n(z)$. It is given by
\begin{equation}\label{eq:pgamma}
P_\gamma(\ell)=\frac{9\Om^2 H_0^4}{4c^4}
\int_0^{z_\mathrm{lim}} \frac{\dd z}{H(z)} \, (1+z)^2 W^2(z) P_\mat^{\mathrm {nl}} (k,z),
\end{equation}
where $k=\ell/S_K(\chi(z))$ and
\begin{equation}
W(z)= \int_z^{z_\mathrm{lim}}\dd z' n(z')
\frac{S_K[\chi(z')-\chi(z)]}{S_K[ \chi(z')]}
\end{equation}
defines the lensing efficiency window function, and
$S_K[\chi(z)]$ denotes the comoving angular diameter distance as
function of redshift, which reduces to the comoving radial
distance $\chi(z)$ for spatial flatness $K=0$. The redshift distribution of
sources was determined in Fu \etal (2008) from the photometric redshift
 estimations of Ilbert \etal (2006). We adopt the commonly used fitting
 function of
\begin{equation}
n(z)=N(z/z_\mathrm{s})^\alpha\exp\left[-(z/z_\mathrm{s})^\beta\right]
\end{equation}
 with best-fit parameters $\alpha=1.47$, $\beta=2.15$, and
 $z_\mathrm{s}=0.90$, and a
 normalization constant $N$ that was determined by integrating up to
 $z_\mathrm{lim}\simeq 6$.

The shear power spectrum is integrated over the
appropriate window function to obtain the mass variance in apertures
$\theta$ \cite{map},
\begin{equation}
\Map{\theta}=\frac{288}{\pi\theta^4}\int\frac{\dd\ell}{\ell^3}J^2_4(\ell\theta)P_\gamma(\ell),
\end{equation}
where $J_4$ is the 4th-order Bessel function of the first kind. 

All quantities relying on the background evolution depend on the
equation-of-state of the various matter components. 
The neutrino equation-of-state is defined as $w_\nu(m_\nu,z)=
P(m_\nu,T(z))/\rho(m_\nu,T(z))$. We compute the pressure P and the energy
density $\rho$ at points of a grid $(m_\nu,T(z))$, using the fact that the
distribution of neutrinos in the phase space is a Fermi-Dirac distribution.
The result is well approximated by the fitting function
\begin{equation}\label{eq:nz}
w_\nu(m_\nu,z)=\frac{1}{3}\left[1+\left(\frac{m_\nu}{(1+z)\,0.058\,\mathrm{eV} }\right)^{a}\right]^{-b},
\end{equation}
where $a=1.652$ and $b=0.561$, and is shown
in Fig.~\ref{fig:wmz+deltaMap2} (left panel). The
transition to the non-relativistic regime occurs earlier for
heavier particles, at the redshift given by Eq.~(\ref{eq:ztrans})
and indicated in the figure. The lensing efficiency window shows the
redshift range probed by our data. At the upper end of this range,
$w_\nu$ is far from zero, especially for low masses. There is thus
a degeneracy between the characteristic redshift of the sources
and the neutrino mass, which is independent of the well-known
redshift-mass degeneracy defined by the amplitude of the cosmic shear
signal.

The total matter power spectrum in Eq.~(\ref{eq:pgamma}) is
computed following Hannestad \etal (2006), which is also similar to the
description adopted for perturbed quintessence fields \cite{csquintessence}.
By assuming that the neutrino overdensities remain always in the linear regime while
CDM and baryons grow non-linearly, the total matter power spectrum is given by
\begin{equation}
\label{eq:Pnl}
P_\mat^\nonlin(k,z)=\left[ f_\nu
\sqrt{P_\nu^\lin(k,z)}+(f_\cdm+f_\baryons)\sqrt{P_{\cdm+\baryons}^\nonlin(k,z)}\right]^2,
\end{equation}
where $f_i\equiv\Omega_i/\Om$ with $i=\{\nu,\cdm,\baryons\}$ are
the density fractions of each matter field over the total matter
density parameter $\Om=\Ocdm+\Ob+\On$. The linear power spectra
$P_i^\lin(k,z)$ are computed using the Boltzmann code
CAMB \footnote{\tt http://camb.info}, to take
into account properly the scale-dependent growth of the neutrino
perturbations. The output is computed for a grid of redshifts
linearly spanning the range $0<z\leq 4$ and then
spline-interpolated.
The non-linear spectrum $P_{\cdm+\baryons}^\nonlin$ is computed with the
{\sc halofit} mapping \cite{halofit}. The Peacock \& Dodds mapping \cite {PD} is unsuitable
for this study, because it uses a scale-independent growth factor, which cannot
be defined in the presence of neutrinos.

The right panel of Fig.~\ref{fig:wmz+deltaMap2} shows the aperture
mass dispersion in the presence of massive and massless neutrinos,
computed from the total matter power spectrum,  for the same $\Om$.
 The models are similar on large scales where the
neutrinos behave like cold dark matter and show a scale-dependent deviation
on intermediate scales. On small scales, which stay always inside the
free-streaming length, the suppression is constant. Linear perturbation theory
predicts the small-scale suppression to be, $[P^\lin_\mat (f_\nu,z=0) -
  P^\lin_\mat (f_\nu=0,z=0)] /P^\lin_\mat  (f_\nu,z=0) \sim -8f_\nu$
\cite{hudeltap}. For the non-linear power spectrum, the suppression is
higher, around  $-10f_\nu$, as predicted by both numerical simulations
\cite{nunbody} and one-loop corrections \cite{Wong08}. 
For the aperture mass dispersion, the models shown in
Fig.~\ref{fig:wmz+deltaMap2} have a small-scale suppression of $\sim
-5f_\nu$. 

\subsection{Likelihood analysis}\label{sec2b}

\subsubsection{Cosmic shear alone}

The statistical analysis evaluates a mixed dark matter scenario
with nine parameters, including the reduced Hubble parameter $h$,
the density of baryons $\omega_\baryons\equiv\Ob\hh$, cold dark
matter, $\omega_\cdm\equiv\Ocdm\hh$, and massive neutrinos,
$\omega_\nu\equiv\On\hh$, the effective number of relativistic
degrees of freedom $\Neff$, the optical depth to reionization
$\tau$, the primordial spectral index $\ns$, the rms of the
matter perturbations extrapolated to redshift $z=0$
and filtered over the $8h^{-1}$~Mpc scale, $\sigma_8$, and the
mean redshift $z_\mathrm{m}$ of the source galaxies.  The results
will be marginalized over a mean redshift range
$0.78<z_\mathrm{m}<1.0$, corresponding to the $2\sigma$ interval
of the redshift distribution found in Fu \etal (2008). The
neutrino physical density parameter relates to the total neutrino
mass as $\On\hh=\summnu/93$~eV, and we assume that the three neutrino
masses are degenerate.

In a very conservative approach, we shall work in a
Friedmann-Lema\^itre-Robertson-Walker (FLRW) cosmology with
cosmological constant as dark energy, with energy density fixed
assuming spatial flatness.  More generally, in this framework, dark
energy is a fluid-like component that is dominant at low redshift and
responsible for the recent  accelerated expansion. A quintessence
field, eventually allowing for non-minimal couplings
\citep{Wetterich1995, uzan99, TocchiniValentiniAmendola2002}, provides the most general
alternative that might alleviate the coincidence problem
\cite[e.g.,][for a review]{QuintessenceReview}. Alternatively, it is
possible to account for the global dynamics without advocating
such an additional field, by  considering back-reaction effects of
structure formation \citep{DEbackreaction, Kolb06, Wiltshire07}, rejecting the
Copernican principle \cite{CopernicanPrinciple}, or invoking
theories of gravitation that differ from general relativity on large
scales \cite[see][for a review]{DEclassification}. Our results do
not exclude these possibilities, which require new, consistent
analyses to be evaluated properly \cite[see e.g.,][for a
weak-lensing analysis of quintessence]{quintlens}.

The log-likelihood is defined as usual to be
\begin{equation}\label{eq:logL}
\log\mathcal{L}=-\frac{1}{2}\sum_{ij}\Delta\Map{\theta_i}(\mathsf{Cov}^{-1})_{ij}\Delta\Map{\theta_j},
\end{equation}
where $\Delta\Map{\theta_i}$ is the difference between the observed and the theoretical
values of the aperture mass variance computed at angles $1 \arcmin \leq \theta_i \leq
230\arcmin$ and $\mathsf{Cov}$ is the corresponding covariance matrix.

We sample the likelihood over a regular grid of the parameter space, for a
 total of $\sim 10^7$ models.  
The domain of the grid is : $\omega_\baryons \in[0.0186,0.0249]$, $\omega_\cdm
 \in[0.110,0.152]$, $\omega_\nu \in[0,0.042]$, $h \in[0.62,0.80]$, $\Neff
 \in[2,4]$, $\tau \in[0,0.225]$, $\ns \in[0.87,1.02]$, $\sigma_8 \in[0.65,0.92]$.

In this work we use the aperture mass vector and covariance matrix obtained in
Fu \etal (2008), which was measured over 57 square degrees (35 square degrees of
effective area) in three connected patches of the sky, using $2.5 \times 10^6$
galaxies with magnitudes $21.5 < i_\mathrm{AB} < 24.5$.
The covariance matrix includes shape noise in the shear
estimator, non-Gaussian cosmic variance, and the residual B-mode. 
The shear-measurement pipeline was tested on STEP2
simulations \cite{step2} and shown to be slightly biased, underestimating
the shear on average by 2\% \cite{Fu}. Moreover, this analysis showed that highly anisotropic point-spread-functions
(PSF) may introduce a spurious constant shear. The aperture mass statistic,
being computed with a compensated filter, is unaffect by a constant shear, which is the main reason for choosing this statistic for the present analysis.
In contrast, a constant shear affects top-hat two-point statistics. The impact of
redshift-dependent additive and multiplicative shear biases, which are present
in all KSB-based shear estimators, on cosmological
constraints is studied in Semboloni \etal (2008).

\subsubsection{Joint analysis}

In the joint analysis, we adopt the {\em importance
sampling} technique \citep{hastings70, lewisbridle},
 adding CFHTLS-T0003 cosmic shear data to two Monte
Carlo Markov chains available in the {\sc Lambda} archive\footnote{\tt
  http://lambda.gsfc.nasa.gov/} for the mixed dark-matter scenario.
One of the chains was computed for CMB temperature and polarization
anisotropies derived from WMAP-5yr \cite{wmap5dunkley} data.
 The other chain used a combination of these data with
baryon acoustic oscillations from SDSS and 2dFGRS
\cite{baodata}, and Type Ia supernovae from the ``Gold'' sample
\cite{sngold} and SNLS \cite{snls}. 

Importance sampling consists of estimating a target
distribution (a joint posterior distribution in our case) by sampling
an auxiliary distribution. We assume that the public Monte Carlo Markov
chains provided by the WMAP team have enough resolution at the region of 
 parameter space intersecting the cosmic shear posterior distribution. 
This is a reasonable assumption since they are the dominant contribution to the
joint constraints. They are thus useful auxiliary distributions and are
biased distributions of the target distributions.
To obtain the two unbiased joint distributions (WMAP+CFHTLS and
WMAP+BAO+SNe+CFHTLS), we scale the chains by 
multiplying the weight of each chain element by its likelihood with respect to
the cosmic shear data.

This method allows us to accelerate the computation 
compared to sampling directly the joint posterior with a grid or a
Markov chain, since it requires only the computation of the cosmic shear
likelihood for each model of the chains. 
Each model of the chains has a constant $\Neff=3.04$, hence in the joint
analysis, we consider only 7+1 physically independent parameters, i.e., the
remaining 7 cosmological parameters used in the cosmic shear analysis and the
lensing sources redshift.

\section{Results}\label{sec3}

\subsection{CFHTLS-T0003 alone}\label{sec3a}

\begin{figure}[t]
\centering
\includegraphics[width=7cm]{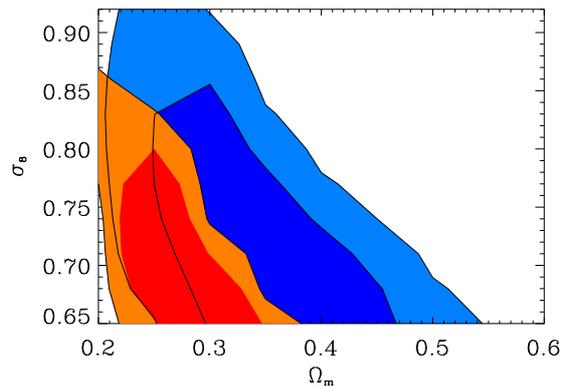}
\vspace{0.4cm}
\caption{Confidence contours (68\% and 95\%) from the aperture-mass dispersion between 1\arcmin and 230\arcmin, for $\On=0$ (red, smaller) and marginalized over $\omega_\nu$ (blue, larger). Both cases are marginalized over the source redshift distribution and the remaining cosmological parameters.}
\label{fig:nufu}
\end{figure}

We now discuss the analysis of cosmic-shear data alone. Figure~\ref{fig:nufu} shows the marginalized confidence contours in the $(\sigma_8,\Omega_\mat)$ plane. The degeneracy direction in the case of massless neutrinos is fitted by
\begin{equation}
\sigma_8\,(\Om/0.25)^{0.72}=0.744 \pm 0.052.
\end{equation}
This value was found by marginalizing over $h$ but for fixed $\omega_\baryons$
and $\ns$ to be more directly comparable to the result of Fu \etal
(2008) for the same data. Our result is slightly lower, but within $1\sigma$.
  There are several differences between the
two analyses that account for this difference. In particular, since we are
interested in combining our results with CMB constraints, we explored a
narrower grid in $(\omega_\baryons,\omega_\cdm,\omega_\nu)$, corresponding to
a grid interval of $\Om \in[0.2,0.6]$. The boundary effects slightly increase the
curvature of the contour, which is most accurately described with a larger exponent power-law. 
The comparison of both results is particularly sensitive to this choice
because here the confidence intervals are computed as volumes of the posterior
distribution containing a certain fraction of the total probability, whereas
in Fu \etal (2008) the results are obtained from the values of
$\Delta\chi^2$. The theoretical modelling also differs. Here
we use Eq.~(\ref{eq:Pnl}), while in Fu \etal (2008) the power spectrum was
computed from the Eisenstein \& Hu transfer function \cite{ehutransfer}.
 
In the presence of massive neutrinos, the contours are shifted towards the right. 
This follows from the $\Ocdm-\On$ degeneracy, and shows that an increase in
$\On$ is compensated by an increase in $\Ocdm$, confirming that the data are in 
the sub-free-streaming-length regime, where the effect of the neutrinos
opposes that of the cold dark matter. The contours also broaden, since many
more preferred models exist now. For example, on angular scales of between
10\arcmin and 100\arcmin,  $\Map{\theta}$ of the best-fit models of the full
grid and the massless neutrino sub-grid, differ by less than 5\%. 
As the contours broaden, their shape remains unchanged, implying that the CDM
effect on the cosmic shear signal is more significant than that of the neutrinos. The degeneracy
 direction in the case of massive neutrinos is now fitted by
\begin{equation}
\sigma_8\,(\Om/0.35)^{0.64}=0.711 \pm 0.076 \,,
\end{equation}
where we chose a higher pivot $\Om$.

\begin{figure*}[t]
\vspace{0.4cm}
\centering
\hspace{-0.4cm}
\includegraphics[width=6cm]{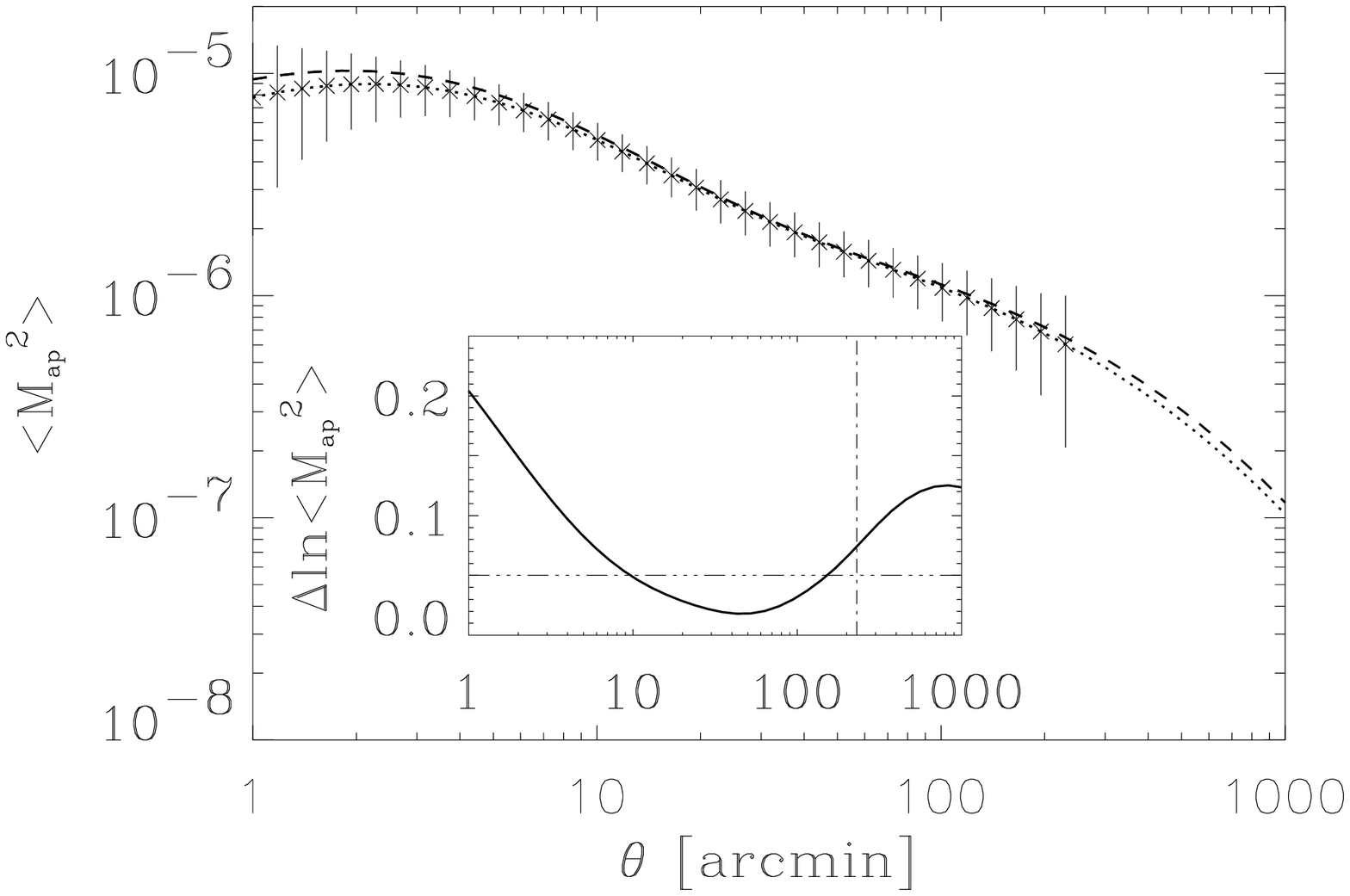}
\includegraphics[width=6cm]{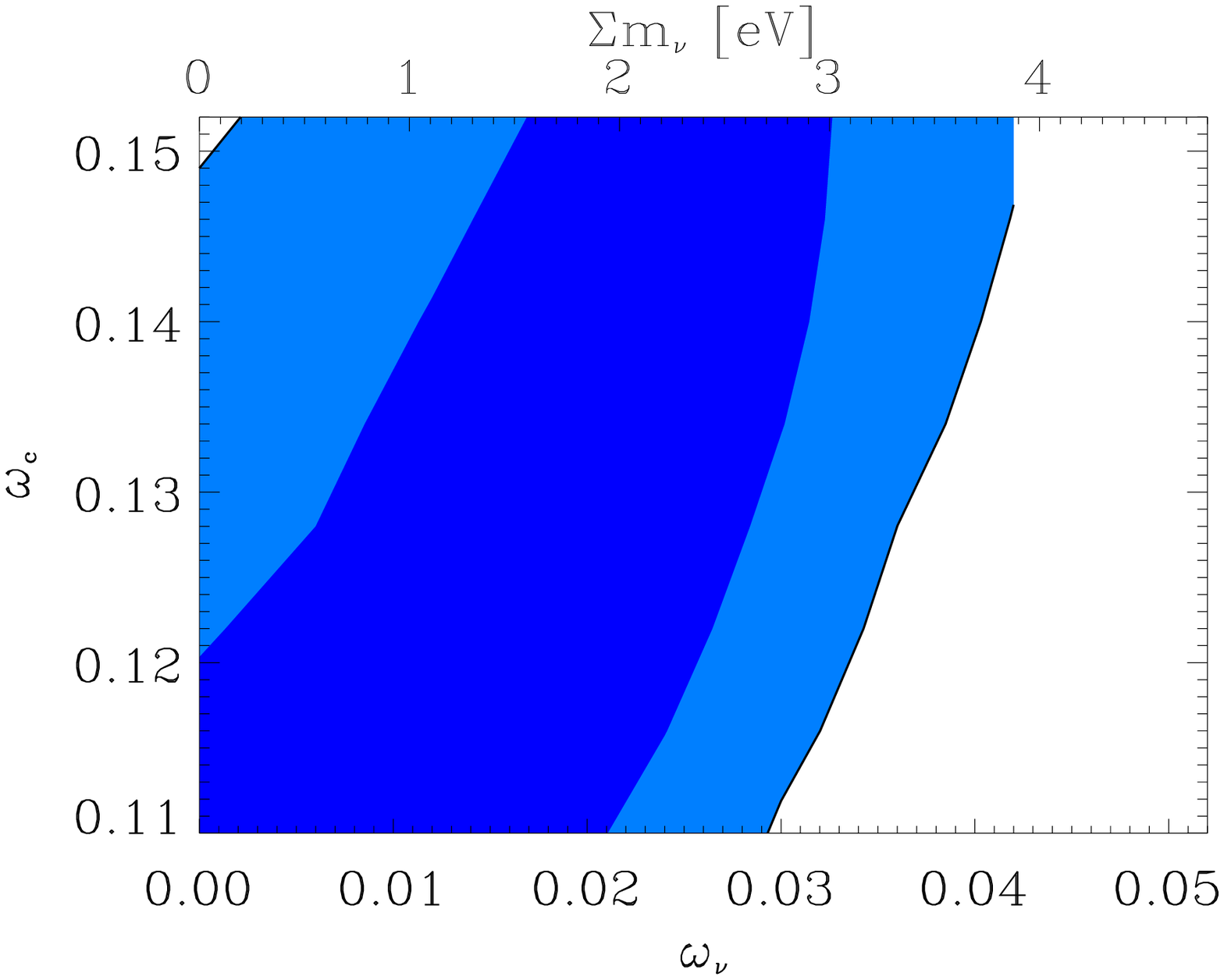}
\includegraphics[width=6cm]{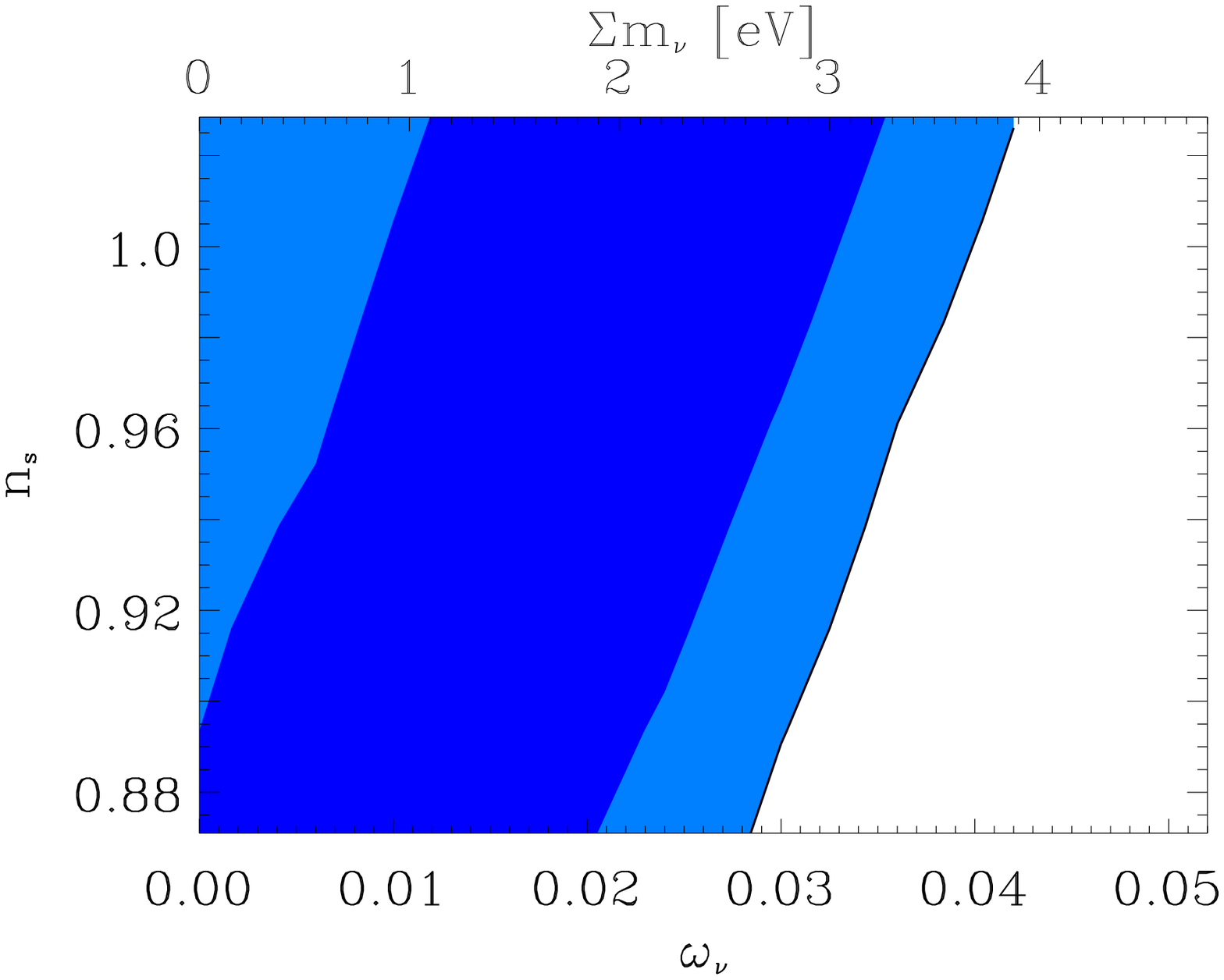}
\vspace{0.4cm}
\caption{
{\it Left:}  Best-fit models found for massive ($m_\nu=0.53$~eV, $\Om=0.44$; dashed)  and massless ($m_\nu=0$~eV, $\Om=0.30$; dotted) neutrino cases. Error bars are from CFHTLS-T0003. The models are extended beyond the data limit to illustrate the model separation at large scales. The normalized difference between the models is shown in the inset, where the horizontal line marks a 5\% model-separation and the vertical line shows the data limit.
{\it Middle and Right:}
Confidence contours (68\% and 95\%) from the aperture-mass variance, in the
$(\omega_\nu,\omega_\cdm)$ and $(\omega_\nu,\ns)$ planes, marginalized over
the hidden parameters.}
\label{fig:wlcont}
\end{figure*}

Figure~\ref{fig:wlcont} (left panel) shows the massless and massive neutrino best-fit models, ($m_\nu=0$~eV, $\Om=0.30$) and ($m_\nu=0.53$~eV, $\Om=0.44$), respectively. They have the same values of $\sigma_8$ and $h$ and differ in $\ns$. The difference between the two models, shown in the inset normalized by the massless neutrino model, increases on larger scales with the approach to the
free-streaming-length.
This indicates that combining cosmic shear measurements at sub-
and super-free-streaming-scales, might break the $\Ocdm-\On$
degeneracy. Shown explicitly in
Fig.~\ref{fig:wlcont} (middle panel), this degeneracy, as well as the
similarly-oriented $\sigma_8-\On$ correlation (not shown), confirm
that an increasing neutrino density decreases the cosmic shear
signal. However, the tilting effect of the neutrino density on the
matter power spectrum around the free-streaming scale may mimic
the dependence on $\ns$, especially if the pivot scale $k$ used is
similar to the free-streaming scale $k^*_\mathrm{fs} \sim 0.01
h$~Mpc$^{-1}$. Therefore, the combination of sub- and super-free-streaming data
may not have enough information to additionally break the $\ns-\On$ degeneracy. This
degeneracy is shown in the right panel of Fig.~\ref{fig:wlcont}.

To explore the benefit of a broader range of scales, we construct a fiducial
 model consisting of an extension of our best-fit
 massless neutrino model to $\theta=20\degr$ (also shown in
 Fig.~\ref{fig:wlcont}, left panel), where the flat-sky approximation
 remains valid. We computed a new covariance matrix using the Schneider \etal
 (2002) approximation with the Semboloni \etal (2007) non-Gaussian correction
 and WMAP-5yr mean parameter values \cite{wmap5dunkley}. We assumed the
 same sky coverage and galaxy density as in CFHTLS-T0003. In this way we can
 evaluate the benefit of using large scales independent of a gain due to
 better statistics. We obtain constraints similar to the ones obtained for
 the data up to 230\arcmin, the extension to $\theta=20\degr$ only adding a 
small number of
 independent points. Since the cosmic shear signal on large
 scales becomes very small, better statistics are needed in attempting
 to break the $\Ocdm-\On$ degeneracy with cosmic shear data alone.

Using either CFHTLS-T0003 data or its extension to
$\theta=20\degr$, the marginalized upper bound on the
neutrino mass is,  
\begin{equation}
\sum m_\nu < 3.3\,\mbox{eV (95\% C.L.)}.
\label{res33}
\end{equation}
This constraint, while very loose when compared with the combined constraints
mentioned in Sect.~2, is comparable to the ones obtained from SDSS or 2dF
galaxy redshift surveys alone (see e.g., Kristiansen \etal 2007, who find
$\summnu \lesssim 5.2 $~eV).

We note that the 95\% C.L. contours in Fig.~\ref{fig:nufu} do not close
inside the range of the parameters probed. This implies that our grid limits
are an effectively strong prior, in particular implying $\Om \in[0.2,0.6]$
 and $\sigma_8 \in[0.65,0.92]$. In other words, the constraint in Eq.~(\ref{res33})
 includes a marginalisation over an arbitrary and relatively narrow range of
 parameters values, and is thus optimistic.
Furthermore, the derived constraint does not take account of
 degeneracies with dark energy.
In more general scenarios with non-vanishing curvature or with
non-cosmological-constant dark energy, both
growth of structure and distances, as functions of redshift, depend on both
the dark-energy density parameter $\Omega_\de$ and the equation-of-state
$w_\de$. To effectively break the
$\On-\Om$ degeneracy, it will thus be necessary to
use a measure that is not degenerate in $\Omega_\de$ and $w_\de$.
One possibility is to exploit the lens efficiency at
different redshift bins with tomography \cite[e.g.,][]{denutom}.
However, even with tomography, some direct neutrino-dark energy degeneracy
will remain due to the
transition epoch from matter to dark-energy domination
\cite{denu}.

Regarding the effective number of relativistic species, cosmic shear is
sensitive to it via the change in the matter-radiation equality. A higher
$\Neff$ implies a longer radiation-dominated era, and thus a more efficient
suppression of growth. Hence, the relative amplitude of small and large scales
changes and the shear power spectrum tilts. The effect is however weak and
we find an essentially flat likelihood in the probed range of $\Neff \in
[2,4]$, meaning that the data do not constrain this parameter.

\begin{figure*}[t]
\vspace{0.4cm}
\centering
\hspace{-0.4cm}
\includegraphics[width=6cm]{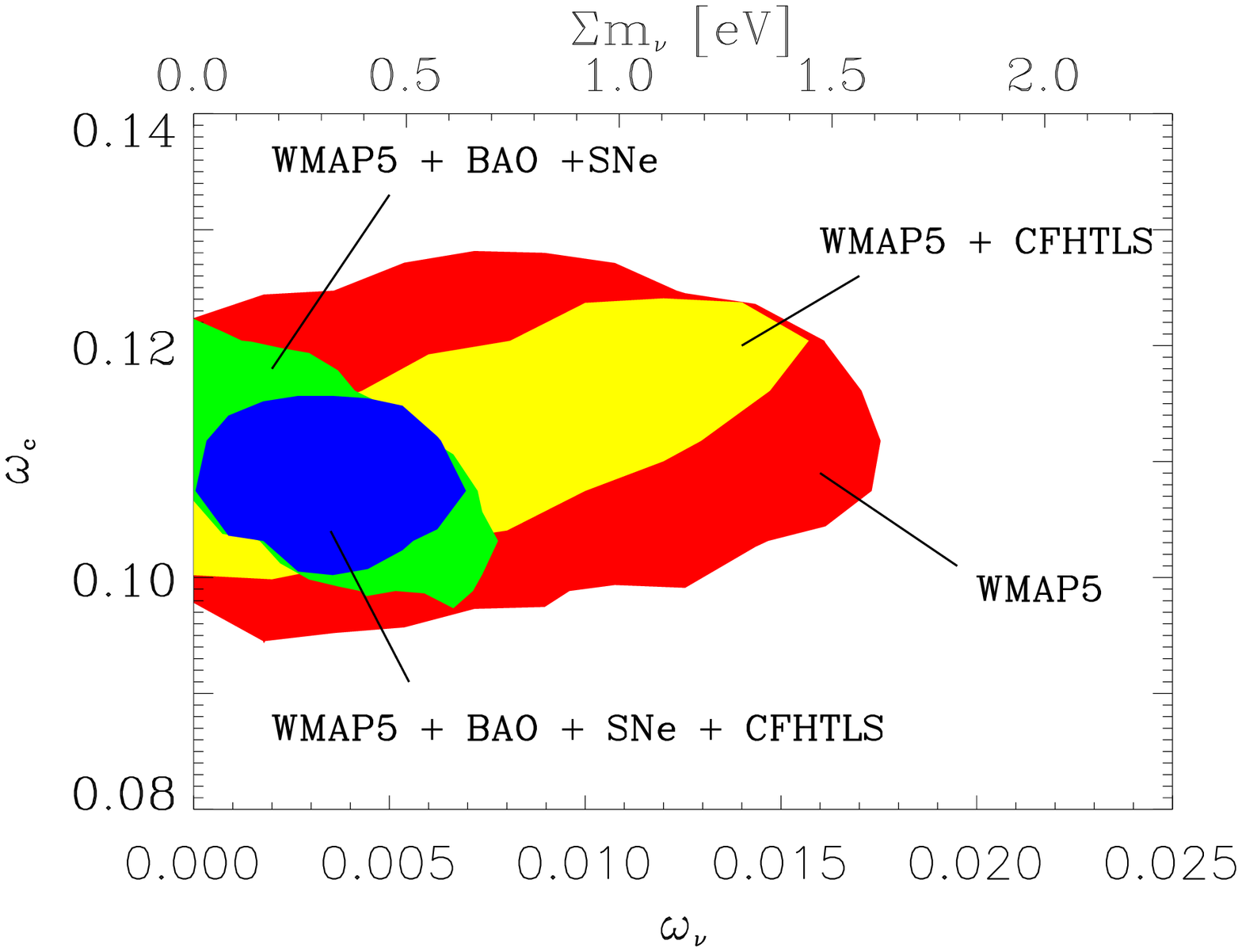}
\includegraphics[width=6cm]{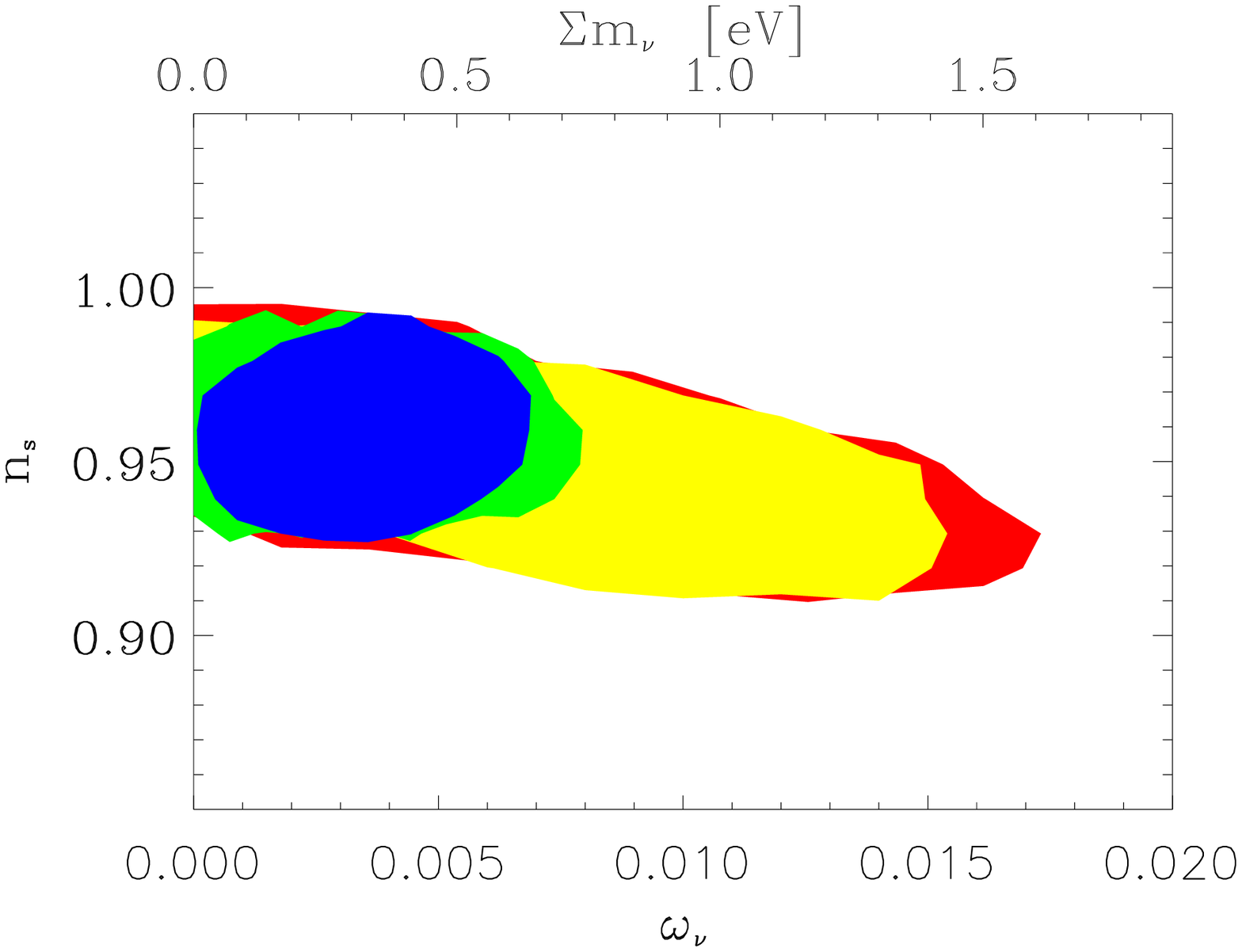}
\includegraphics[width=6cm]{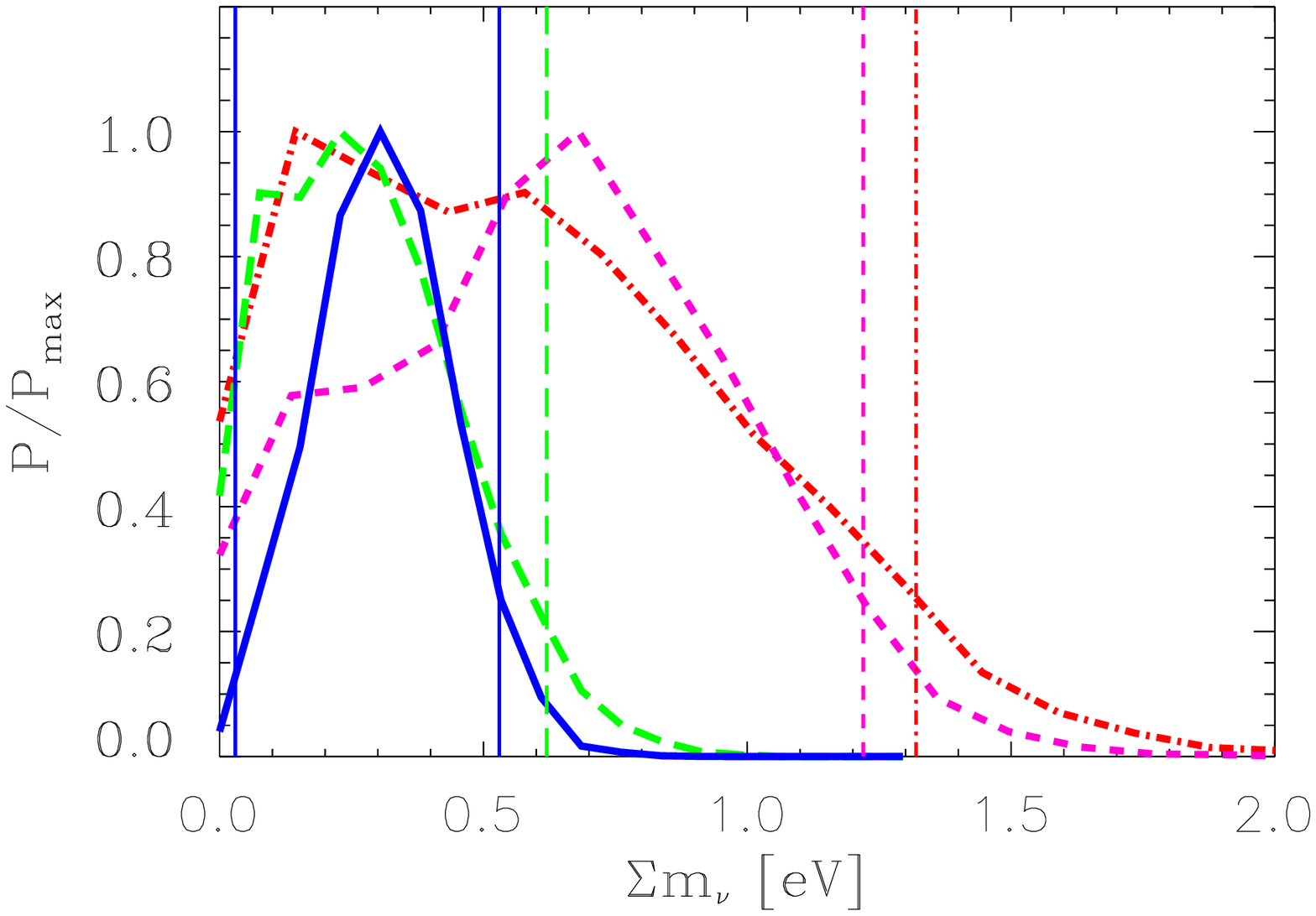}
\vspace{0.4cm}
\caption{Joint analysis.
  {\it Left and Middle:} Contours at 95\% C.L. of  $(\omega_\nu,\omega_\cdm)$ and $(\omega_\nu,\ns)$
  for the four cases considered~:  WMAP5 alone (largest contour, red), WMAP5+CFHTLS (second largest, yellow),
  WMAP5+BAO+SNe (second smallest, green), WMAP5+BAO+SNe+CFHTLS (smallest contour, blue).
  {\it Right:} One-dimensional marginals for $\summnu$ showing the 95\% C.L
  with vertical lines, for the same four cases~: WMAP5 (dash-dot,
  red);  WMAP5+CFHTLS (dash, pink) ;
  WMAP5+BAO+SNe (long dash, green); and WMAP5+BAO+SNe+CFHTLS (solid, blue). For
  the latter a lower bound was also found).}
\label{fig:joint}
\end{figure*}

\subsection{Joint analysis}\label{sec3b}

 We now explore the parameter
space by adding CFHTLS-T0003 cosmic shear data to each of the Monte Carlo Markov
chains defined in Sect. 2.2.2, aiming to determine the gain
achieved by including lensing data.

The main independent piece of information provided by cosmic shear is the
power spectrum on small scales. As discussed above, it has a
distinctive scaling with both neutrino and CDM densities, while 
these two parameters for CMB are instead weakly correlated, as seen in
Fig.~\ref{fig:joint} (left panel, largest contour). This allows us
to obtain a narrow, joint constraint, approximatively in the same
direction found for cosmic shear alone in Fig.~\ref{fig:wlcont} (middle
panel). 
The direction orthogonal to the contour indicates the combination of parameters to which
cosmic shear is mostly sensitive to. It is well approximated by the linear relation 
 $\omega_\cdm - \omega_\nu$ and the WMAP5 + CFHTLS constraint on this combination of
parameters is,
\begin{equation}
\omega_\cdm - \omega_\nu = 0.106 \pm 0.006\; (1\sigma).
\end{equation}
The figure-of-merit (FoM) quantifies the error ellipses of
correlated parameters and was introduced for the parameters of the dark
energy equation of state \cite{detf}.
It is defined as the inverse area of the ellipse approximating the $p\%$
confidence level contour, centered on the contour centroid, i.e.,
\begin{equation}
\mathrm{FoM}^{-1}=\Delta\chi^2(p)\pi\,(\sigma_{11}\sigma_{22}-\sigma_{12}^2)^{1/2},
\end{equation}
where $\sigma_{ii}$ is the variance of the parameter $i$.
We consider 2$\sigma$ contours, for which $\Delta\chi^2(95.4\%)=6.17$, and
obtain ${\rm FoM} \simeq 19$ for the WMAP5 contour and ${\rm FoM} \simeq
44$ for the combined (WMAP5 + CFHTLS) case, corresponding to a gain of a
factor of 2.3.
The two parameters $\omega_\cdm$ and $\omega_\nu$ are, however, strongly
correlated, exhibiting a narrow, long combined contour and the effective gain
in the $\omega_\nu$ variance is of only 1.2.

The introduction of distance measurements (BAO+SNe) is useful, since
its combination with WMAP5 data (also shown
in Fig.~\ref{fig:joint}, left panel) defines a narrow region in the
$(\omega_\cdm,\omega_\nu)$ plane that complements that of the
WMAP5+CFHTLS data.
 Both cases thus probe orthogonal combinations of the two parameters and,
 furthermore, in a
 way that is independent of the spectral index, since this is already well
 determined by the CMB data alone (as shown in Fig.~\ref{fig:joint}, middle panel).
 Their combination thus has the same properties we were looking for when discussing the possibility of
 using sub- and super-free-streaming scales from cosmic shear alone, along
 with the need for an independent measure of $\ns$. This means that the
 combination of the two cases (WMAP5 + BAO + SNe + CFHTLS) breaks the remaining
 degeneracy, as is clear from  Fig.~\ref{fig:joint} (left panel, smaller contour) and from
 the fact that the combined contour has very small correlation ($\sigma_{12}<
 0.1\%$, for un-normalised parameters $\omega_c$ and $\omega_\nu$).

Figure~\ref{fig:joint} (right panel) shows the one-dimensional marginalized
probability distributions of the neutrino masses in the four cases. 
The upper 95\% confidence levels are $\sum m_\nu < 1.32$~eV (WMAP5), 
$\sum m_\nu < 1.22$~eV (WMAP5 + CFHTLS) and  $\sum m_\nu < 0.62$~eV (WMAP5 +
BAO + SNe). 
For the full combination (WMAP5 + BAO + SNe + CFHTLS), we find that
\begin{equation}
0.03\,\mbox{eV}\, < \sum m_\nu < 0.54\,\mbox{eV (95\% C.L.)},
\end{equation}
with mean 0.31~eV. Interestingly, we obtain a lower bound, and thus
 a preference for massive neutrinos at the $2\sigma$ level.

It is worth to point out that the lower bound on $\sum m_\nu$ is 
strongly dependent on the position of the WMAP5 + CFHTLS contour in the 
$(\omega_\nu,\omega_\cdm)$ plane (see Fig.~\ref{fig:joint}, left panel).
Any systematic effect that underestimates
 the cosmic-shear signal would shift the $(\Om,\sigma_8)$ contour
 towards the bottom-left corner of Fig.~\ref{fig:nufu} with a consequent shift
 in the $(\omega_\nu,\omega_\cdm)$ contour towards the bottom-right corner of 
 Fig.~\ref{fig:joint} (left panel).
 The WMAP5 + BAO + SNe + CFHTLS contour would thus descend along the WMAP5 + BAO + SNe
one producing a positive lower bound for the neutrino mass. A similar effect
appeared in Allen \etal (2003), which also showed
a preference for a non-zero neutrino mass from a combination of CMB, galaxy
clustering and X-ray cluster data; that result was in fact caused by an underestimation
of $\sigma_8$,  as pointed out by Seljak \etal (2005).
Similarly, the joint constraint depends crucially on BAO data. Komatsu
  \etal (2008) pointed out a tension between SNe and BAO results when the joint
SDSS + 2dFGRS BAO sample is used. If BAO data overestimates the total matter
density, then an unbiased WMAP5 + BAO + SNe contour in the 
$(\omega_\nu,\omega_\cdm)$ plane (Fig.~\ref{fig:joint}, left panel) will be
shifted towards the bottom-left corner, weakening the preference for massive neutrinos. 

\subsection{Robustness of the constraints}\label{sec3c}

\begin{figure*}[t]
\vspace{0.4cm}
\centering
\hspace{-0.2cm}
\includegraphics[width=5.9cm]{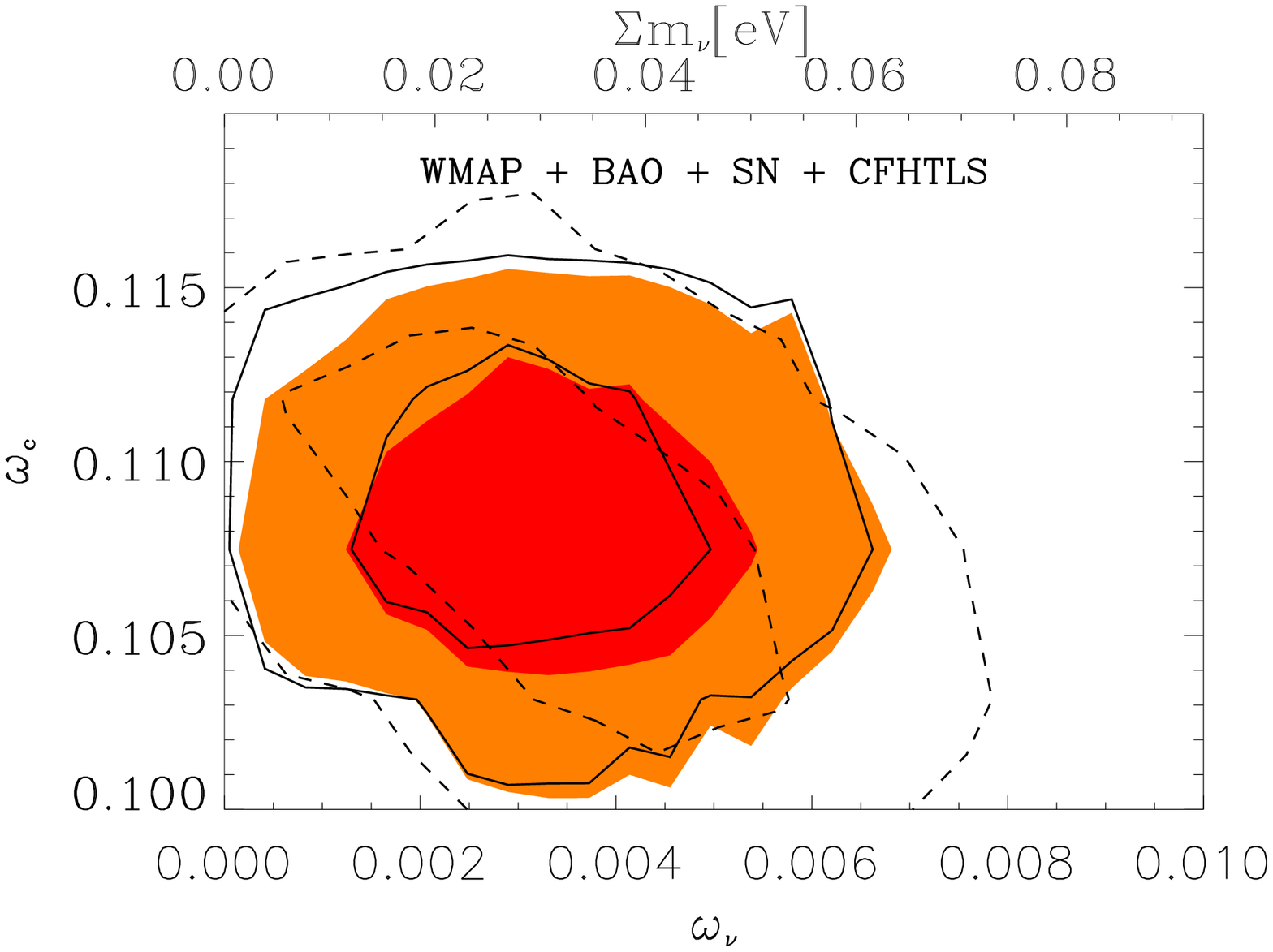}
\hspace{0.1cm}
\includegraphics[width=5.9cm]{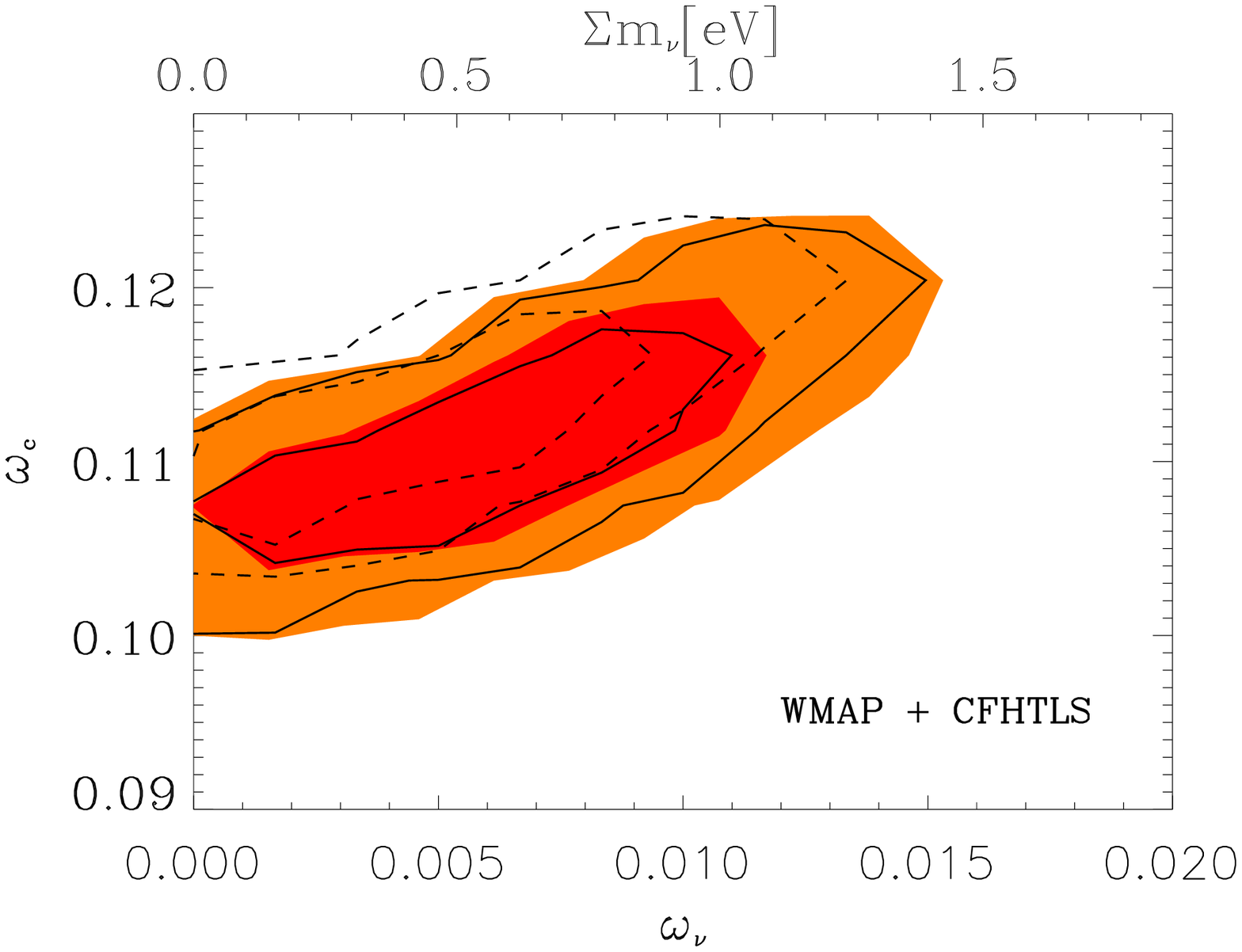}
\hspace{0.1cm}
\includegraphics[width=5.9cm]{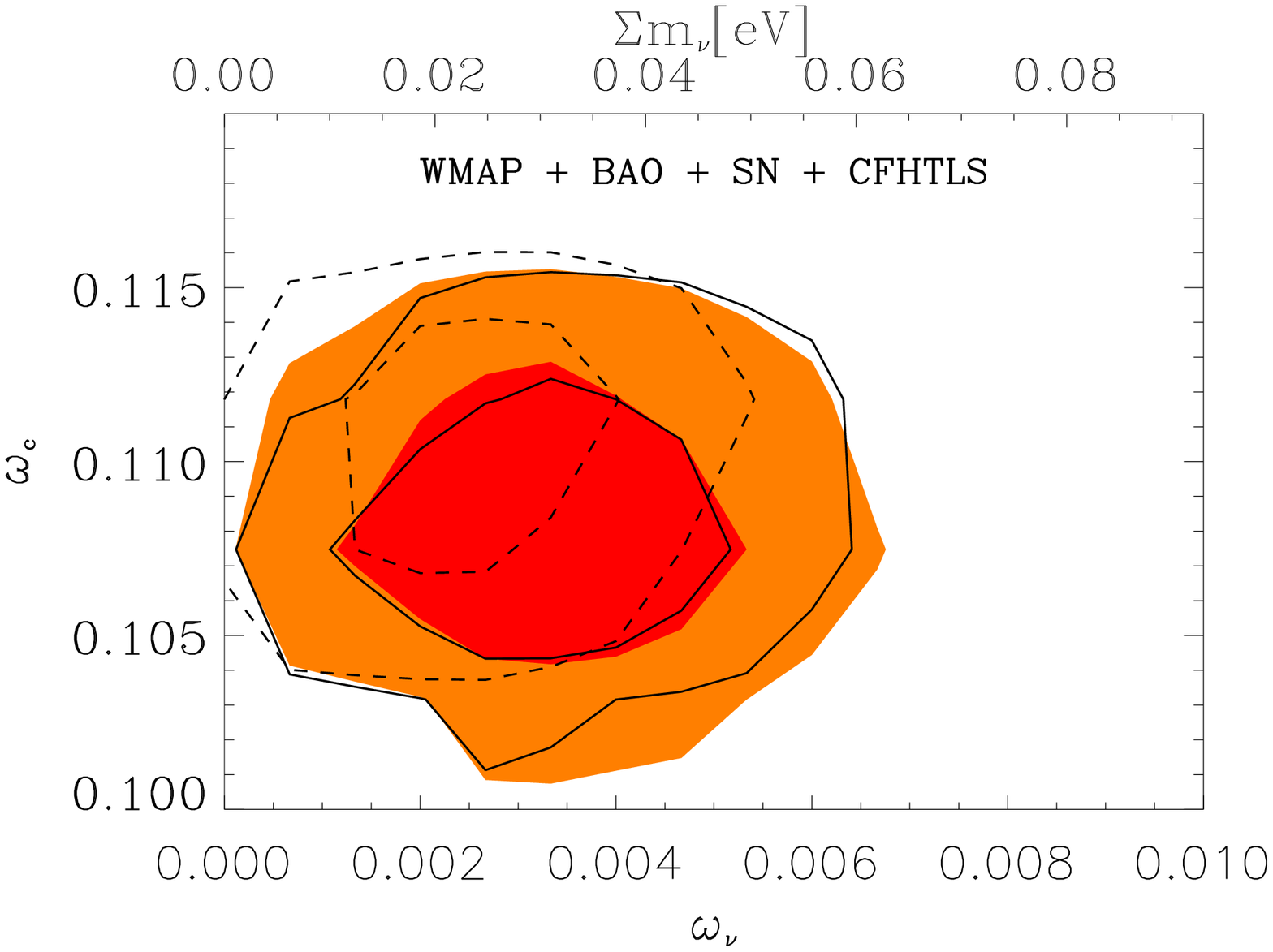}
\vspace{0.4cm}
\caption{Tests of the robustness of the result. Marginalized joint constraints
  (68\% and 95\% C.L.) obtained using the actual data
  (filled contours) and various assumptions (open contours). 
{\em Left:} Impact of systematics, assuming a 4\% underestimation (solid) or
  marginalizing over a $\pm 25\%$ calibration bias (dashed).
{\em Middle and Right:} Forecasts using a synthetic covariance matrix and the
  CFHTLS vector data (solid) or the same covariance matrix with a
  WMAP5 fiducial model (dashed).
}
\label{fig:testWLdata}
\end{figure*}

We check here the robustness of our results against systematics of the cosmic
shear data, that have not been included in the analysis so
far. For this purpose, we first assume the data vector is biased as a
consequence of an underestimation in the shear, as indicated by the STEP simulations.
We assume a 2\% redshift-independent bias in the shape measurements, which
translates into a scale-independent 4\% error in the data vector, and repeat the joint analysis. 
The resulting combined contour shifts upwards along the WMAP5+BAO+SNe
degeneracy in agreement with the discussion in Sect. 3.2. 
As seen in Fig.~\ref{fig:testWLdata} (left panel, solid lines), this
 small bias is enough for the evidence of massless neutrinos to be lost and we
 obtain at the 95\% C.L.~: 
\begin{equation}
\sum m_\nu<0.53~eV.
\end{equation}

Other sources of errors that can affect cosmic-shear cosmological constraints
are: contamination from intrinsic alignments, in particular shear-shape
correlations \cite{hirataseljak} for no precise theoretical modeling yet exists; uncertainties in the photometric
redshifts; and extra PSF residuals not predicted by the STEP simulations.
Modelling of some of the CFHTLS-T0003 systematics is included in the analysis
of Kilbinger \etal (2008). Here we consider the rougher approach of marginalizing
 over a scale-independent calibration factor, accounting for all possible
 sources of biases, with the goal of finding a threshold of contamination
 above which the CFHTLS data do not improve the combined constraints.

 We find a threshold of 25\% for the aperture mass dispersion data. The
 corresponding combined contour in the $(\omega_\nu,\omega_\cdm)$ plane 
is shown in Fig.~\ref{fig:testWLdata} (left panel, dashed lines).
The contour is now elongated and similar to the WMAP+BAO+SNe one of
Fig.~\ref{fig:joint} (left panel).
If the data contains systematics of amplitude smaller than 25\% of the signal,
 we can quote conservative combined constraints
by marginalizing over the amplitude of the systematics. The final 95\% C.L. for the
joint analysis will be
intermediate between the most optimistic case (no bias) of 0.03~eV $< \sum m_\nu<0.54$~eV
and the worst case (no contribution from lensing) of $\sum m_\nu<0.62$~eV. For example, in the
case of systematics of 10\%, the conservative constraints are $\sum m_\nu<0.58$~eV. 
These marginalized constraints, not including systematics
 with a scale- or a redshift-dependence, are approximative since massive
 neutrinos affect both the amplitude and the shape of the cosmic shear correlations.
 Only on smaller scales ($\lesssim 10\arcmin \,$, see
 e.g. Fig.~\ref{fig:wmz+deltaMap2} right panel) is the suppression
 scale-independent.

We can also assess the robustness of the result by comparing it with a forecast.
For this, we compute a systematics-free
cosmic-shear covariance matrix for our survey size using
the results of Schneider \etal (2002) and Semboloni \etal (2007).
 This covariance matrix differs from the one we have used so far, 
 which was computed by taking into account the true galaxy positions and weights. 
 The results do not change when we replace the covariance matrices,
 as shown in Fig.~\ref{fig:testWLdata} (middle and right panels, solid lines). 
To ensure that the forecast is completely
 independent of the data, we redo it by further replacing the data vector by a
 fiducial model \cite[the WMAP5 mean,][]{wmap5dunkley}. This time we obtain
 a shift in the contour location (Fig.~\ref{fig:testWLdata}, middle and right
 panels, dashed lines). Again the evidence for massless neutrinos is lost,
 this time due to the higher $\omega_\cdm$ value of the WMAP5
fiducial model. This behaviour mimicks a correction for
an eventual overestimation of the source redshifts. Indeed, if the effective
redshift is lower, the models will have a higher value of $\Om$ for the same amplitude
$\sigma_8$, shifting the combined $\omega_\cdm - \omega_\nu$ contour up.  
 It likewise mimicks a correction for an underestimation of the shear measurements.
The combined constraint on the neutrino mass, when using the WMAP5 fiducial model,
strengthens to $\sum m_\nu<0.44$~eV. We note this is also the result that
would be obtained if the data were corrected for a negative calibration bias of 20\%.

In summary, the possible systematics do not change the angle between the
$\omega_\cdm-\omega_\nu$ degeneracy in the WMAP5 + CFHTLS  and WMAP5 + BAO +
SNe cases, only shift or broaden the combined contour position and size.
Eventual corrections for an underestimation of the data vector or for an
overestimation of the source redshifts would produce both a tighter upper
bound and a looser lower bound on the neutrino mass. 

The matter power spectrum also contributes as a source
of uncertainties. The prescription used, Eq.~(\ref{eq:Pnl}), does not 
consider clustering of neutrinos on CDM structures, which occurs
when the neutrino thermal velocity drops below the velocity dispersion, $v$,
 of forming clusters; for instance, it takes place at $z\sim 2.3$ for $v\sim
 1000$~km/s and $m_\nu\sim 0.5$~eV. 
The corresponding neutrino halo profile is flatter in the centre than a pure
CDM Navarro-Franck-White \cite{nuprofile}. 
Including it in the 1-halo term of the halo model, Abazajian \etal (2005)
showed that it decreases the non-linear matter power spectrum. Accordingly, in this
work the term $P_{\cdm+\baryons}^\nonlin$ 
is expected to be overestimated by $\sim1\%$ on scales around
$k=0.5\,h$~Mpc$^{-1}$ for $m_\nu \sim 0.5$~eV.
Alternatively, results from perturbation theory with neutrinos \cite{Wong08} also indicate 
an overestimation of the matter power spectrum, as already mentioned in
Sect.~2.1.

The non-linear power spectrum is computed with the {\sc halofit}, which declares a 3\%
uncertainty on scales $k<10$~h/Mpc at $z<3$ \cite{halofit}, which are thus of the
same order as the supposed shear measurement bias. 
 This uncertainty is effectively both redshift- and scale-independent over the data redshift and scale ranges.
Marginalizing over a 5\% uncertainty, the corresponding joint constraint is $\sum m_\nu<0.56$~eV. 
Furthermore, the {\sc halofit} does not take account of the effects of
cooling baryons and hot intra-cluster baryons, which are degenerate with the
neutrinos, affecting the power spectrum by a few percent
 on small scales \citep{psbaryons1, psbaryons2, psbaryons3}.

\section{Conclusions}\label{sec4}

We have investigated the potential of cosmic shear to constrain the mass of
neutrinos. In the sub-free-streaming regime, the constraining power
originates, for a fixed density of baryons, in the tendency of relativistic
(hot) dark matter particles to escape from collapsed regions. Therefore
 additional amounts of CDM are needed to produce the same cosmic-shear distortion, shifting
the $\Om-\sigma_8$ degeneracy towards larger values of $\Om$ with
respect to the analysis without massive neutrinos
(Fig.~\ref{fig:nufu}), producing a $\omega_\cdm-\omega_\nu$
degeneracy favoring higher amounts of CDM for higher amounts of
massive neutrinos (Fig.~\ref{fig:wlcont}). The analysis of
CFHTLS-T0003 data alone yields a loose constraint of $\sum
m_\nu<3.3$~eV at the 95\% C.L., for our particular choice of priors.

We have explored larger angular scales using a synthetic data vector extended
to $20\degr$ to explore the possibility of breaking the
$\omega_\cdm-\omega_\nu$ degeneracy using only cosmic shear.
 This analysis did not predict an improvement in
the results, which would require higher signal-to-noise ratio at
large scales. 

The introduction of CFHTLS-T0003 data in a WMAP5+BAO+SNe analysis provides
 an interesting combination, breaking the
$\omega_\cdm-\omega_\nu$ degeneracy present in that analysis
(Fig.~\ref{fig:joint} , left panel). The joint analysis 
yields 0.03~eV $< \sum m_\nu<0.54$~eV at the 95\%
confidence level, marginally excluding massless neutrinos
(Fig.~\ref{fig:joint}, right panel). The preference for massive neutrinos is
lost when the data is corrected for a possible underestimation of the shear
signal, which simulations indicate may be around 4\%. In this case, the result
shifts to $\sum m_\nu<0.53$~eV. The CFHTLS data also shows some disagreement
with the result of WMAP5. Indeed, if a cosmic-shear fiducial WMAP5 model was used,
the expected combined WMAP5+BAO+SNe+CFHTLS constraint would be tighter: $\sum m_\nu<0.44$~eV. 
 Finally, we showed that CFHTLS-T0003 data do not improve the WMAP5+BAO+SNe constraints on
 the neutrino mass if they contain a bias larger than 25\%.

 After submission of the manuscript, a similar study was submitted by Ichiki
\etal (2008), also using CFHTLS-T0003 cosmic shear data to constrain the
neutrino mass. Both analyses take a similar approach, the main differences
being: they use the cosmic shear correlation function, which is more affected by
systematics, as opposed to the aperture-mass dispersion; the 
non-Gaussian shear covariance matrix is also computed in different ways, as is the
likelihood analysis. Their neutrino mass constraints from cosmic shear only
are weaker, which is consistent with the fact they impose weaker priors, and the
results for a WMAP5+BAO+SNe+CFHTLS joint analysis are very similar.

 Besides the neutrino mass, we also considered the effective number of relativistic degrees of freedom, $\Neff$.
 CMB anisotropies are by far more sensitive than cosmic shear and CFHTLS-T0003
 data produced an almost flat likelihood with respect to this parameter. This
 prevented us from using the $\sigma_8-\Neff$ degeneracy, caused by a higher
 $\Neff$ delaying the matter-radiation equivalence and the corresponding
 defreezing of matter perturbations, which have less time to grow.

In summary, the CFHTLS cosmic-shear has already sufficient statistical precision for the accuracy of the results to be affected by
systematics. The CFHTLS lensing systematics collaboration is currently
undertaking a very detailed and lengthy analysis of those systematics.
The statistical precision is however still insufficient to allow us to 
improve current constraints on the neutrino mass.
 Future ground-based and space-borne observations \cite[such as KIDS,
 Pan-STARRS, DES, LSST, SNAP-L, JDEM,  or EUCLID surveys; see][]{ESAESOreport},
 with better statistics, larger scales, and also the use of tomographic techniques, will allow to 
 perform more elaborate analyses,
 for instance enabling to relax the assumption of degeneracy between mass
 eigenstates.

 From the theoretical point-of-view, a more suitable computation of the
 matter power spectrum in the non-linear regime will be mandatory, to take
 more careful account of the effects of massive neutrinos without
 relying on mappings based on $\Lambda$CDM $N$-body simulations. This might be
 achievable using either standard perturbation theory \cite[e.g.,][]{nupt} or a 
resummation scheme \citep[see e.g.,][]{ptr1, ptr2}.
 It will be
 interesting, eventually, to investigate the degeneracy with the low-redshift
 physics ($z\lesssim 20$), such as reionization and dark energy, relevant if
 $m_\nu\lesssim0.01$~eV.
 On this mass scale, the normal/inverted hierarchy of mass states can also be
 distinguished.
 More elaborate models allowing for an interaction between massive neutrinos
 and a quintessence field \cite{mQ2}, or in which
 mass-varying neutrinos behave as a negative pressure fluid, which could then
 be the origin of the cosmic acceleration \cite{mQ1}, may then also be considered.

\begin{acknowledgements}
We are thankful to Yannick Mellier for supporting this project and to
Elisabetta Semboloni for many discussions.
We thank Karim Benabed for help with computation in the early
stage of this work and the Terapix group for computational facilities. 
We acknowledge the CFHTLS lensing systematics collaboration for helpful discussions. 
We are grateful to Peter Schneider and Jens Roediger
for careful readings of the manuscript.
We acknowledge use of CAMB and of the LAMBDA archive.
 IT and LF are supported by the
European Commission Programme 6-th framework, Marie Curie Training
and Research Network  ``DUEL'', contract number
MRTN-CT-2006-036133. MK is funded by the CNRS/ANR research grant
``ECOSSTAT'', contract number ANR-05-BLAN-0283-04.
CS thanks IAP for hospitality.

\end{acknowledgements}


\end{document}